\let\Xdocument\document
\documentclass{iau}
\let\document\Xdocument

\usepackage{amsmath}
\usepackage{graphicx}
\usepackage{multirow}
\usepackage[flushleft]{threeparttable}

\begin{document}

\lefttitle{S. Das et al.}
\righttitle{A multi-wavelength analysis of BL Her stars}

\jnlPage{1}{8}
\jnlDoiYr{2023}
\doival{10.1017/xxxxx}

\aopheadtitle{Proceedings IAU Symposium 376}
\editors{Richard de Grijs, Patricia Whitelock and M\'arcio Catelan, eds.}

\title{A multi-wavelength analysis of BL Her stars:\\  Models versus Observations}

\author{S. Das$^1$, L. Moln\'ar$^1$, S. M. Kanbur$^2$, M. Joyce$^1$,
  A. Bhardwaj$^3$, H. P. Singh$^4$,\\ M. Marconi$^3$, V. Ripepi$^3$
  and R. Smolec$^5$ }

\affiliation{$^1$Konkoly Observatory, Research Centre for Astronomy and Earth Sciences, E\"{o}tv\"{o}s Lor\'and Research Network (ELKH), Konkoly-Thege Mikl\'os \'ut 15-17, H-1121, Budapest, Hungary \\
$^2$Department of Physics, State University of New York Oswego, Oswego, NY 13126, USA\\
$^3$INAF-Osservatorio Astronomico di Capodimonte, Salita Moiariello 16, 80131, Naples, Italy\\
$^4$Department of Physics \& Astrophysics, University of Delhi, Delhi 110007, India\\
$^5$Nicolaus Copernicus Astronomical Center, Polish Academy of Sciences, Bartycka 18, PL-00-716 Warsaw, Poland
             }

\begin{abstract}
We present new theoretical period--luminosity (PL) and period--radius
(PR) relations at multiple wavelengths (Johnson--Cousins--Glass and
{\sl Gaia} passbands) for a fine grid of BL~Herculis models
computed using {\sc mesa-rsp}. The non-linear models were computed for periods typical
of BL~Her stars, i.e. $1\leq P ({\rm days}) \leq4$, covering a wide
range of input parameters: metallicity ($-$2.0 dex $\leq$ [Fe/H]
$\leq$ 0.0 dex), stellar mass (0.5--0.8 M$_{\odot}$), luminosity
(50--300 L$_{\odot}$) and effective temperature (full extent of the
instability strip; in steps of 50K). We investigate the impact of four
sets of convection parameters on multi-wavelength
properties. Most empirical relations match well with theoretical
relations from the BL~Her models computed using the four sets of
convection parameters. No significant metallicity effects are seen in
the PR relations. Another important result from our grid of BL~Her
models is that it supports combining PL relations of RR Lyrae and
Type~II Cepheids together as an alternative to classical Cepheids for
the extragalactic distance scale calibration.
\end{abstract}


\maketitle

\section{Introduction}

Classical pulsators, in particular RR~Lyrae stars, classical and
Type~II Cepheids are extremely important astrophysical objects; they
are commonly used to estimate extragalactic distances, thanks to their
well-defined period--luminosity (PL) relations, especially at longer
wavelengths \citep[for a review, see][]{2020JApA...41...23B}. However,
although the effect of metallicity on the PL relations of RR~Lyrae
stars and classical Cepheids is minimised at near-infrared
wavelengths, the precise calibration of PL-metallicity (PLZ) relations
is one of the most important current topics of research in stellar
variability studies \citep[for example,
  see][]{2021MNRAS.508.4047R,2022A&A...659A.167R,2022ApJ...939...89B,
  2022ApJS..262...25D}. Type~II Cepheids, on the other hand, are shown
to exhibit little to no metallicity dependence on the PL relations
\citep{2011MNRAS.413..223M, 2017A&A...604A..29G,
  2021MNRAS.501..875D}. Type~II Cepheids are located in the classical
instability strip of the Hertzsprung--Russell diagram. They have
luminosities brighter than RR~Lyrae but fainter than classical
Cepheids. This makes them useful tracers, especially in regions with
faint RR~Lyrae and/or scarce classical Cepheids \citep[for a review,
  see][]{2002PASP..114..689W}.

Type~II Cepheids are further subclassified on the basis of their
pulsational periods \citep{2018AcA....68...89S} into BL~Her stars ($1
\lesssim P \textrm{(days)}\lesssim 4$), W~Vir stars ($4 \lesssim P
\textrm{(days)}\lesssim 20$) and RV~Tau stars ($P \gtrsim 20$
days). An additional class includes the peculiar W~Vir (pW~Vir) stars,
which are brighter and bluer than the normal W~Vir stars
\citep{2008AcA....58..293S}. In this study, we analyse only the
shortest-period Type~II Cepheids, i.e., BL~Her stars, from both
theoretical and empirical points of view. This is because
\textsc{mesa-rsp} \citep{2019ApJS..243...10P} can reliably model
BL~Her stars but not the longer-period Type~II Cepheids (W~Vir and
RV~Tau stars) owing to their highly non-adiabatic nature. As alluded
to already, both empirical \citep{2006MNRAS.370.1979M,
  2009MNRAS.397..933M, 2011MNRAS.413..223M, 2017A&A...604A..29G} and
theoretical \citep{2007A&A...471..893D, 2021MNRAS.501..875D} studies
provide evidence of largely negligible effects of metallicity on the
PL relations of BL~Her stars and Type~II Cepheids, in
general. However, recent results from \cite{2022ApJ...927...89W}
indicate a significant effect of metallicity on the PL relations of
field Type~II Cepheids at near-infrared wavelengths, albeit with the
caveat that their sample size was rather small, containing only 23
Type~II Cepheids. If confirmed with more observational data using
high-resolution spectroscopy and multiband photometry, this
discrepancy between cluster and field Type~II Cepheids could
potentially turn into an open problem.

In this study, we use the state-of-the-art radial stellar pulsation
code \textsc{mesa-rsp} to compute a fine grid of BL~Her models using
the four sets of convection parameters, derive new theoretical PL,
period--Wesenheit (PW) and period--radius (PR) relations at multiple
wavelengths (Johnson--Cousins--Glass $UBVRIJHKLL'M$ and {\sl Gaia}
$GG_{\rm BP}G_{\rm RP}$ bands) and thereby study the effects of
metallicity and convection parameters on these relations. We also
compare our theoretical results with empirical results of BL~Her stars
in the LMC from {\sl Gaia} Data Release 3
\citep[DR3;][]{2019A&A...625A..14R} and from
\citet{2017A&A...604A..29G} using OGLE data.

\section{Data and Methodology}

The computation of BL~Her models using \textsc{mesa-rsp} and the
subsequent final selection of the models used for this analysis have
been described in \cite{2021MNRAS.501..875D}. In brief, the BL~Her
models were computed using \textsc{mesa}~r11701, which uses the
\citet{1986A&A...160..116K} theory of turbulent convection and the
method of stellar pulsation as prescribed by
\citet{2008AcA....58..193S}. The four sets of convection parameters
are from Table~4 of \citet{2019ApJS..243...10P}, each with increasing
complexities: set~A corresponds to the simplest convection model,
set~B has radiative cooling added, set~C included turbulent pressure
and turbulent flux and set~D has all of these effects added
simultaneously.

We begin with the linear computations covering a wide range of input
parameters: metallicity ($-2.0$ dex $\leq$ [Fe/H] $\leq$ 0.0 dex),
stellar mass (0.5--0.8 M$_{\odot}$), luminosity (50--300 L$_{\odot}$)
and effective temperature (full extent of the instability strip; in
steps of 50K) using the four sets of convection parameters, resulting
in a combination of 20,412 models per convection set over the entire
range of input parameters. The linear stability analysis yields the
linear periods and the growth rates of the models in different radial
pulsation modes. This helps us estimate the red and blue edges of the
instability strip as traced by the positive growth rates of the
fundamental-mode BL~Her pulsators \citep[see
  Figure~1;][]{2021MNRAS.501..875D}. We proceed with the non-linear
computations of the models with linear periods between 0.8 and 4.2
days. The non-linear integrations are computed over 4000 pulsation
cycles each and only those models are accepted for the final analysis
that have non-linear periods between 1 and 4 days and satisfy the
condition of full-amplitude stable pulsations \citep[the amplitude of
  radius variation $\delta R$, the pulsation period $P$, and the
  fractional growth of the kinetic energy per pulsation period
  $\Gamma$ do not vary by more than 0.01 over the last 100 cycles of
  the 4000-cycle integrations: see
  Figure~2;][]{2021MNRAS.501..875D}. Models that did not converge
within 4000 cycles were not included in this study. After these
conditions are met, we finally have 3266 BL~Her models computed using
set~A, 2260 models using set~B, 2632 models using set~C and 2122
models using set~D.

\begin{figure}[t]
\centering
\includegraphics[scale = 0.9]{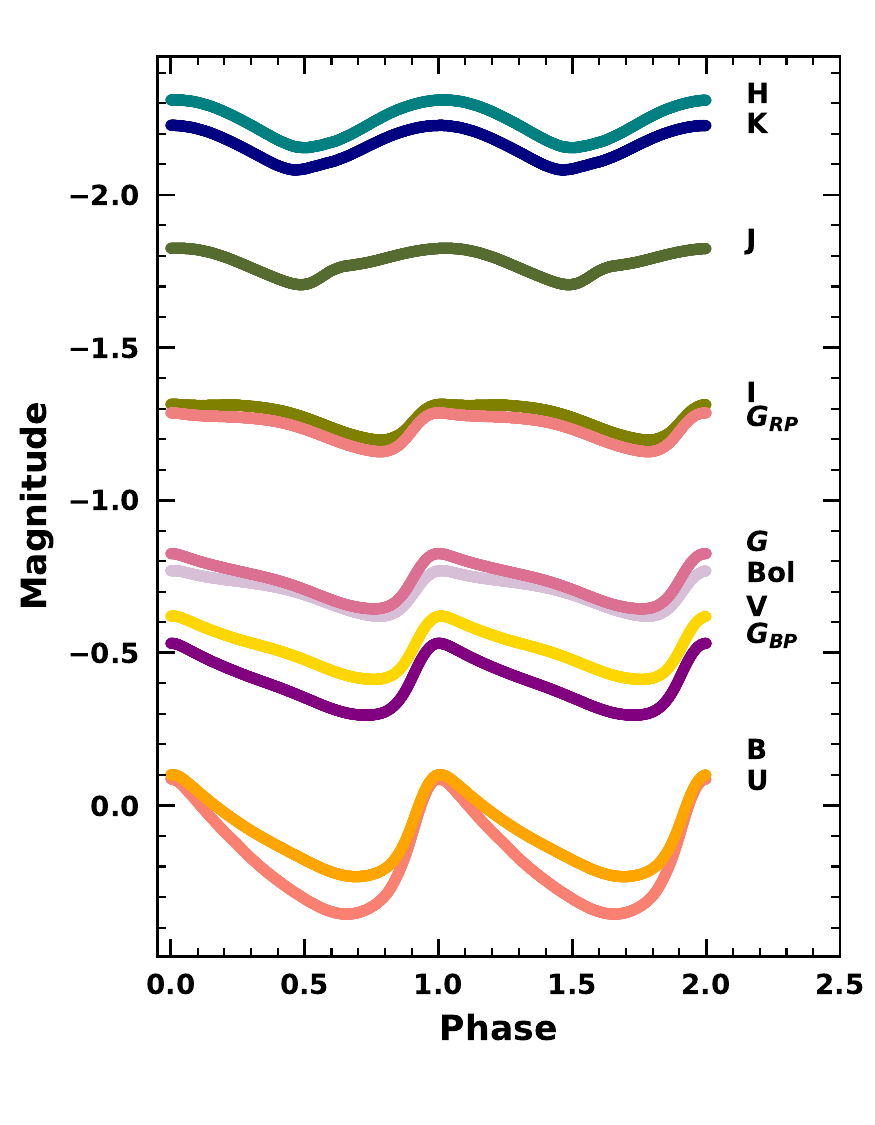}
\caption{Example of multi-wavelength light curves for the BL~Her model
  with input parameters, $Z=0.00135$, $X=0.74806$, $M=0.6$
  M$_{\odot}$, $L=150$ L$_{\odot}$ and $T=5500$ K, based on convective
  parameter set A.}
\label{fig:LC}
\end{figure}

The bolometric luminosities generated as output from the non-linear
computations of \textsc{mesa-rsp} are converted into absolute
bolometric magnitudes and also into absolute magnitudes $M_\lambda$ in
a given passband $\lambda$, using either pre-computed \citep[for
  transformation into Johnson--Cousins--Glass bands
  using][]{1998A&AS..130...65L} or user-provided (for transformation
into {\sl Gaia} passbands using the Packaged Model Grids from MESA
Isochrones \& Stellar Tracks,
MIST\footnote{\url{https://waps.cfa.harvard.edu/MIST/index.html}})
bolometric correction tables. The bolometric correction tables are
defined as a function of the stellar photosphere in terms of $T_{\rm
  eff}$\,(K), $\log g ({\rm cm \; s^{-2}})$, and metallicity,
[M/H]. An example of the light curves in both the bolometric
magnitudes and in the different passbands for a BL~Her model computed
using convection parameter set~A is presented in Figure~\ref{fig:LC}.

The multi-wavelength theoretical light curves are fitted with the
Fourier sine series \citep[for an example, see][]{2018MNRAS.481.2000D}
of the form,
\begin{equation}
m(x) = m_0 + \sum_{k=1}^{N}A_k \sin(2 \pi kx+\phi_k),
\label{eq:fourier}
\end{equation}
\noindent where $x$ is the pulsation phase, $m_0$ the mean
magnitude and $N$ the order of the fit ($N = 20$).

\begin{table*}
\caption{Comparison of the slopes of the PL and PW relations for
  BL~Her stars of the mathematical form $M_\lambda=a\log(P)+b$. $N$ is
  the total number of stars, $T$ represents the observed value of the
  $t$-statistic, and $p(t)$ gives the probability of acceptance of the
  null hypothesis (equal slopes). The bold-faced entries indicate that
  the null hypothesis of the equivalent PL slopes can be rejected.}
\begin{threeparttable}
\centering \scalebox{0.475}{
\begin{tabular}{c c c c c c c c c c c c}
\hline\hline
Band & Source & $a$ & $b$ & $\sigma$ & $N$ & Reference${^\ddagger}$ & Theoretical/ & \multicolumn{4}{c}{($T$, $p(t)$) w.r.t.}\\
& & & & & & & Empirical & Set A & Set B & Set C & Set D\\
\hline
Bolometric & $Z_{\rm all}$ (Set A) &$-$1.799$\pm$0.028&$-$0.181$\pm$0.01&0.253&3266& TW & Theoretical & ... & ... & ... & ...\\
Bolometric & $Z_{\rm all}$ (Set B) &$-$1.581$\pm$0.03&$-$0.18$\pm$0.01&0.231&2260& TW & Theoretical & \textbf{(5.245,0.0)}& ... & ... & ...\\
Bolometric & $Z_{\rm all}$ (Set C) &$-$1.693$\pm$0.031&$-$0.094$\pm$0.011&0.256&2632& TW & Theoretical & \textbf{(2.532,0.006)} & \textbf{(2.609,0.005)} & ... & ...\\
Bolometric & $Z_{\rm all}$ (Set D) &$-$1.559$\pm$0.032&$-$0.103$\pm$0.011&0.246&2122& TW & Theoretical & \textbf{(5.625,0.0)} & (0.512,0.304) & \textbf{(3.051,0.001)} & ...\\
Bolometric & LMC & $-$1.749$\pm$0.200 & 0.141$\pm$0.051 & 0.274 & 57(4) & G17 & Empirical & (0.248,0.402) & (0.831,0.203) & (0.277,0.391) & (0.938,0.174)\\
\hline
$U$ & $Z_{\rm all}$ (Set A) & $-$0.841$\pm$0.044&0.185$\pm$0.015&0.391&3266& TW & Theoretical & ... & ... & ... & ...\\
$U$ & $Z_{\rm all}$ (Set B) & $-$0.596$\pm$0.046&0.215$\pm$0.015&0.353&2260& TW & Theoretical & \textbf{(3.846,0.0)}& ... & ... & ...\\
$U$ & $Z_{\rm all}$ (Set C) & $-$0.422$\pm$0.051&0.298$\pm$0.018&0.428&2632& TW & Theoretical & \textbf{(6.219,0.0)} & \textbf{(2.512,0.006)} & ... & ...\\
$U$ & $Z_{\rm all}$ (Set D) & $-$0.369$\pm$0.053&0.309$\pm$0.018&0.409&2122& TW & Theoretical & \textbf{(6.851,0.0)} & \textbf{(3.213,0.001)} & (0.722,0.235) & ...\\
\hline
$B$ & $Z_{\rm all}$ (Set A) & $-$1.166$\pm$0.04&0.187$\pm$0.014&0.351&3266& TW & Theoretical & ... & ... & ... & ...\\
$B$ & $Z_{\rm all}$ (Set B) & $-$0.896$\pm$0.041&0.209$\pm$0.013&0.311&2260& TW & Theoretical & \textbf{(4.764,0.0)}& ... & ... & ...\\
$B$ & $Z_{\rm all}$ (Set C) & $-$0.942$\pm$0.043&0.33$\pm$0.015&0.359&2632& TW & Theoretical & \textbf{(3.843,0.0)} & (0.791,0.214) & ... & ...\\
$B$ & $Z_{\rm all}$ (Set D) & $-$0.805$\pm$0.044&0.324$\pm$0.015&0.339&2122& TW & Theoretical & \textbf{(6.1,0.0)} & (1.508,0.066) & \textbf{(2.235,0.013)} & ...\\
\hline
$V$ & $Z_{\rm all}$ (Set A) & $-$1.616$\pm$0.032&$-$0.14$\pm$0.011&0.284&3266& TW & Theoretical & ... & ... & ... & ...\\
$V$ & $Z_{\rm all}$ (Set B) & $-$1.374$\pm$0.034&$-$0.134$\pm$0.011&0.256&2260& TW & Theoretical & \textbf{(5.221,0.0)}& ... & ... & ...\\
$V$ & $Z_{\rm all}$ (Set C) & $-$1.487$\pm$0.034&$-$0.031$\pm$0.012&0.288&2632& TW & Theoretical & \textbf{(2.761,0.003)}& \textbf{(2.343,0.01)} & ... & ...\\
$V$ & $Z_{\rm all}$ (Set D) & $-$1.337$\pm$0.036&$-$0.043$\pm$0.012&0.275&2122& TW & Theoretical & \textbf{(5.829,0.0)} & (0.757,0.225) & \textbf{(3.023,0.001)} & ...\\
\hline
$I$ & $Z_{\rm all}$ (Set A) & $-$2.043$\pm$0.025&$-$0.592$\pm$0.008&0.219&3266& TW & Theoretical & ... & ... & ... & ...\\
$I$ & $Z_{\rm all}$ (Set B) & $-$1.848$\pm$0.027&$-$0.599$\pm$0.009&0.203&2260& TW & Theoretical & \textbf{(5.386,0.0)}& ... & ... & ...\\
$I$ & $Z_{\rm all}$ (Set C) & $-$1.932$\pm$0.027&$$-$$0.534$\pm$0.009&0.226&2632& TW & Theoretical & \textbf{(3.05,0.001)} & \textbf{(2.223,0.013)} & ... & ...\\
$I$ & $Z_{\rm all}$ (Set D) & $-$1.81$\pm$0.028&$-$0.545$\pm$0.01&0.218&2122& TW & Theoretical & \textbf{(6.225,0.0)} & (0.976,0.165) & \textbf{(3.129,0.001)} & ...\\
\hline
$J$ & $Z_{\rm all}$ (Set A) & $-$2.303$\pm$0.021&$-$0.914$\pm$0.007&0.186&3266& TW & Theoretical & ... & ... & ... & ...\\
$J$ & $Z_{\rm all}$ (Set B) & $-$2.131$\pm$0.023&$-$0.928$\pm$0.008&0.177&2260& TW & Theoretical & \textbf{(5.505,0.0)}& ... & ... & ...\\
$J$ & $Z_{\rm all}$ (Set C) & $-$2.239$\pm$0.023&$-$0.877$\pm$0.008&0.19&2632& TW & Theoretical & \textbf{(2.067,0.019)} & \textbf{(3.332,0.0)} & ... & ...\\
$J$ & $Z_{\rm all}$ (Set D) & $-$2.122$\pm$0.024&$-$0.89$\pm$0.008&0.187&2122& TW & Theoretical & \textbf{(5.617,0.0)} & (0.243,0.404) & \textbf{(3.497,0.0)} & ...\\
$J$ & LMC & $-$2.164$\pm$0.240 & 17.131$\pm$0.038 (@0.3)$^{*}$ & 0.25 & 55 & M09 & Empirical & (0.577,0.282) & (0.137,0.446) & (0.311,0.378) & (0.174,0.431)\\
$J$ & LMC & $-$2.294$\pm$0.153 & 15.375$\pm$0.113 (@1.0)$^{\dagger}$ & 0.202 & 55 & B17 & Empirical & (0.058,0.477) & (1.054,0.146) & (0.355,0.361) & (1.111,0.133)\\
\hline
$H$ & $Z_{\rm all}$ (Set A) & $-$2.57$\pm$0.018&$-$1.17$\pm$0.006&0.157&3266& TW & Theoretical & ... & ... & ... & ...\\
$H$ & $Z_{\rm all}$ (Set B) & $-$2.429$\pm$0.02&$-$1.192$\pm$0.007&0.154&2260& TW & Theoretical & \textbf{(5.236,0.0)}& ... & ... & ...\\
$H$ & $Z_{\rm all}$ (Set C) & $-$2.529$\pm$0.019&$-$1.162$\pm$0.007&0.16&2632& TW & Theoretical & (1.568,0.058) & \textbf{(3.587,0.0)} & ... & ...\\
$H$ & $Z_{\rm all}$ (Set D) & $-$2.432$\pm$0.021&$-$1.175$\pm$0.007&0.162&2122& TW & Theoretical & \textbf{(5.048,0.0)} & (0.081,0.468) & \textbf{(3.439,0.0)} & ...\\
$H$ & LMC & $-$2.259$\pm$0.248 &$ $16.857$\pm$0.039 (@0.3)$^{*}$ & 0.26 & 54 & M09 & Empirical & (1.251,0.106) & (0.683,0.247) & (1.086,0.139) & (0.695,0.244)\\
$H$ & LMC & $-$2.088$\pm$0.214 & 15.218$\pm$0.163 (@1.0)$^{\dagger}$ & 0.296 & 52 & B17 & Empirical & \textbf{(2.244,0.012)} & (1.587,0.056) & \textbf{(2.053,0.02)} & (1.6,0.055)\\
\hline
$K$ & $Z_{\rm all}$ (Set A) & $-$2.528$\pm$0.018&$-$1.124$\pm$0.006&0.16&3266& TW & Theoretical & ... & ... & ... & ...\\
$K$ & $Z_{\rm all}$ (Set B) & $-$2.383$\pm$0.021&$-$1.144$\pm$0.007&0.157&2260& TW & Theoretical & \textbf{(5.308,0.0)}& ... & ... & ...\\
$K$ & $Z_{\rm all}$ (Set C) & $-$2.483$\pm$0.02&$-$1.112$\pm$0.007&0.164&2632& TW & Theoretical & \textbf{(1.7,0.045)} & \textbf{(3.526,0.0)} & ... & ...\\
$K$ & $Z_{\rm all}$ (Set D) & $-$2.383$\pm$0.021&$-$1.125$\pm$0.007&0.165&2122& TW & Theoretical & \textbf{(5.194,0.0)} & (0.008,0.497) & \textbf{(3.453,0.0)} & ...\\
$K_{\rm s}$ & LMC & $-$1.992$\pm$0.2$7$8 & 16.733$\pm$0.040 (@0.3)$^{*}$ & 0.26 & 47 & M09 & Empirical& \textbf{(1.924,0.027)} & (1.402,0.081) & \textbf{(1.762,0.039)} & (1.402,0.081)\\
$K_{\rm s}$ & LMC & $-$2.083$\pm$0.154 & 15.162$\pm$0.114 (@1.0)$^{\dagger}$ & 0.262 & 47 & B17 & Empirical & \textbf{(2.87,0.002)} & \textbf{(1.93,0.027)} & \textbf{(2.576,0.005)} & \textbf{(1.93,0.027)}\\
\hline
$G$ & $Z_{\rm all}$ (Set A) & $-$1.76$\pm$0.029&$-$0.263$\pm$0.01&0.262&3266& TW & Theoretical & ... & ... & ... & ...\\
$G$ & $Z_{\rm all}$ (Set B) & $-$1.531$\pm$0.031&$-$0.261$\pm$0.01&0.237&2260& TW & Theoretical & \textbf{(5.352,0.0)}& ... & ... & ...\\
$G$ & $Z_{\rm all}$ (Set C) & $-$1.63$\pm$0.032&$-$0.171$\pm$0.011&0.267&2632& TW & Theoretical & \textbf{(2.996,0.003)} & \textbf{(2.231,0.026)} & ... & ...\\
$G$ & $Z_{\rm all}$ (Set D) & $-$1.49$\pm$0.033&$-$0.183$\pm$0.011&0.255&2122& TW & Theoretical & \textbf{(6.095,0.0)} & (0.895,0.371) & \textbf{(3.049,0.002)} & ...\\
$G$ & LMC &$-$1.248$\pm$0.149&18$.$549$\pm$0.043&0.159&61& {\sl Gaia} DR3 & Empirical & \textbf{(3.373,0.001)} & (1.86,0.063) & \textbf{(2.507,0.012)} & (1.586,0.113)\\
\hline
$G_{\rm BP}$ & $Z_{\rm all}$ (Set A) & $-$1.543$\pm$0.033&$-$0.077$\pm$0.011&0.293&3266& TW & Theoretical & ... & ... & ... & ...\\
$G_{\rm BP}$ & $Z_{\rm all}$ (Set B) & $-$1.296$\pm$0.034&$-$0.068$\pm$0.011&0.262&2260& TW & Theoretical & \textbf{(5.193,0.0)}& ... & ... & ...\\
$G_{\rm BP}$ & $Z_{\rm all}$ (Set C) & $-$1.382$\pm$0.036&0.034$\pm$0.012&0.299&2632& TW & Theoretical & \textbf{(3.33,0.001)} & (1.731,0.084) & ... & ...\\
$G_{\rm BP}$ & $Z_{\rm all}$ (Set D) & $-$1.238$\pm$0.037&0.024$\pm$0.013&0.284&2122& TW & Theoretical & \textbf{(6.178,0.0)} & (1.151,0.25) & \textbf{(2.804,0.005)} & ...\\
$G_{\rm BP}$ & LMC &$-$0.759$\pm$0.254&18.628$\pm$0.074&0.247&55& {\sl Gaia} DR3 & Empirical & \textbf{(3.061,0.002)} & \textbf{(2.095,0.036)} & \textbf{(2.428,0.015)} & (1.866,0.062)\\
\hline
$G_{\rm RP}$ & $Z_{\rm all}$ (Set A) & $-$2.015$\pm$0.025&$-$0.588$\pm$0.009&0.225&3266& TW & Theoretical & ... & ... & ... & ...\\
$G_{\rm RP}$ & $Z_{\rm all}$ (Set B) & $-$1.81$\pm$0.027&$-$0.594$\pm$0.009&0.209&2260& TW & Theoretical & \textbf{(5.503,0.0)}& ... & ... & ...\\
$G_{\rm RP}$ & $Z_{\rm all}$ (Set C) & $-$1.909$\pm$0.028&$-$0.523$\pm$0.01&0.231&2632& TW & Theoretical & \textbf{(2.819,0.005)} & \textbf{(2.565,0.01)} & ... & ...\\
$G_{\rm RP}$ & $Z_{\rm all}$ (Set D) & $-$1.779$\pm$0.029&$-$0.536$\pm$0.01&0.223&2122& TW & Theoretical & \textbf{(6.144,0.0)} & (0.783,0.434) & \textbf{(3.271,0.001)} & ...\\
$G_{\rm RP}$ & LMC & $-$1.782$\pm$0.216&18.041$\pm$0.065&0.219&56& {\sl Gaia} DR3 & Empirical & (1.072,0.284) & (0.129,0.898) & (0.583,0.56) & (0.014,0.989)\\
\hline
$W(G,G_{\rm BP}-G_{\rm RP})$ & $Z_{\rm all}$ (Set A) & $-$2.656$\pm$0.018& $-$1.234$\pm$0.006&0.159&3266& TW & Theoretical & ... & ... & ... & ...\\
$W(G,G_{\rm BP}-G_{\rm RP})$ & $Z_{\rm all}$ (Set B) & $-$2.507$\pm$0.021& $-$1.261$\pm$0.007&0.159&2260& TW & Theoretical & \textbf{(5.432,0.0)}& ... & ... & ...\\
$W(G,G_{\rm BP}-G_{\rm RP})$ & $Z_{\rm all}$ (Set C) & $-$2.633$\pm$0.019& $-$1.231$\pm$0.007&0.163&2632& TW & Theoretical & (0.887,0.375) & \textbf{(4.412,0.0)} & ... & ...\\
$W(G,G_{\rm BP}-G_{\rm RP})$ & $Z_{\rm all}$ (Set D) & $-$2.517$\pm$0.021& $-$1.247$\pm$0.007&0.165&2122& TW & Theoretical & \textbf{(4.98,0.0)} & (0.341,0.733) & \textbf{(3.994,0.0)} & ...\\
$W(G,G_{\rm BP}-G_{\rm RP})$ & LMC &$-$2.362$\pm$0.205&17.312$\pm$0.061&0.206&58& {\sl Gaia} DR3 & Empirical & (1.429,0.153) & (0.704,0.482) & (1.316,0.188) & (0.752,0.452)\\
\hline
\multicolumn{12}{c}{RR~Lyrae PL relations compared with BL~Her PL relations}\\
\hline
$R$ & $Z_{\rm all}$ &$-$1.756$\pm$0.077&$-$0.114$\pm$0.014&0.196&226& M15 & Theoretical & (1.19,0.117) & (1.485,0.069) & (0.208,0.418) & \textbf{(1.882,0.03)}\\
$I$ & $Z_{\rm all}$ &$-$1.973$\pm$0.068&$-$0.415$\pm$0.013&0.175&226& M15 & Theoretical & (0.966,0.167) & \textbf{(1.709,0.044)} & (0.561,0.287) & \textbf{(2.217,0.013)}\\
$J$ & $Z_{\rm all}$ &$-$2.245$\pm$0.06&$-$0.778$\pm$0.011&0.155&226& M15 & Theoretical & (0.902,0.184) & \textbf{(1.769,0.039)} & (0.098,0.461) & \$textbf{(1.898,0.029)}\\
$H$ & $Z_{\rm all}$ &$-$2.206$\pm$0.118&$-$1.043$\pm$0.022&0.302&226& M15 & Theoretical & \textbf{(3.056,0.001)} & \textbf{(1.867,0.031)} & \textbf{(2.708,0.003)} & \textbf{(1.889,0.03)}\\
$K$ & $Z_{\rm all}$ &$-$2.514$\pm$0.057&$-$1.11$\pm$0.011&0.147&226& M15 & Theoretical & (0.24,0.405) & \textbf{(2.149,0.016)} & (0.507,0.306) & \textbf{(2.149,0.016)}\\
\hline
\end{tabular}}
\begin{tablenotes}
	\small
	\item ${^\ddagger}$ TW = This work; M09 = \cite{2009MNRAS.397..933M}; M15 = \cite{2015ApJ...808...50M}; B17 = \cite{2017AJ....153..154B}; \\
\: G17 = \cite{2017A&A...604A..29G}
	\item $^{*}$ Zero point at $\log(P)=0.3$    
	\item $^{\dagger}$ Zero point at $\log(P)=1.0$     
\end{tablenotes}
\label{tab:PL}
\end{threeparttable}
\end{table*}

\begin{figure}[th!]
\centering
\includegraphics[scale = 0.8]{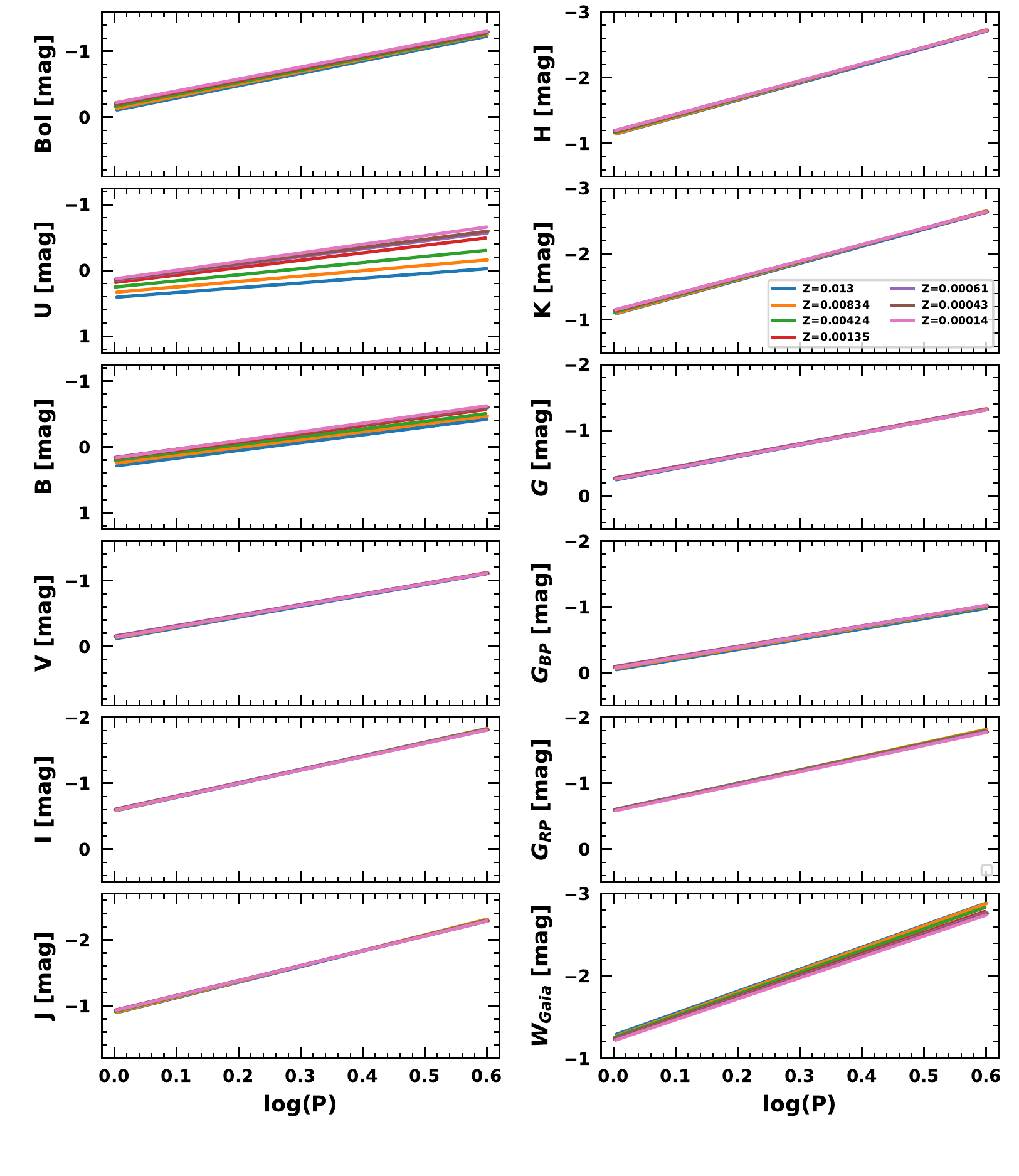}
\caption{PL relations of the BL~Her models with different chemical
  compositions at different wavelengths for the convective parameter
  set~A. The $y$-axis scale is same (2.5~mag) in each panel to
  facilitate a relative comparison. The other convection parameter
  sets (B, C and D) show similar PL relations as a function of
  metallicity and wavelength.}
\label{fig:PL_Zlambda}
\end{figure}

\begin{figure}[th!]
\centering
\includegraphics[scale = 0.9]{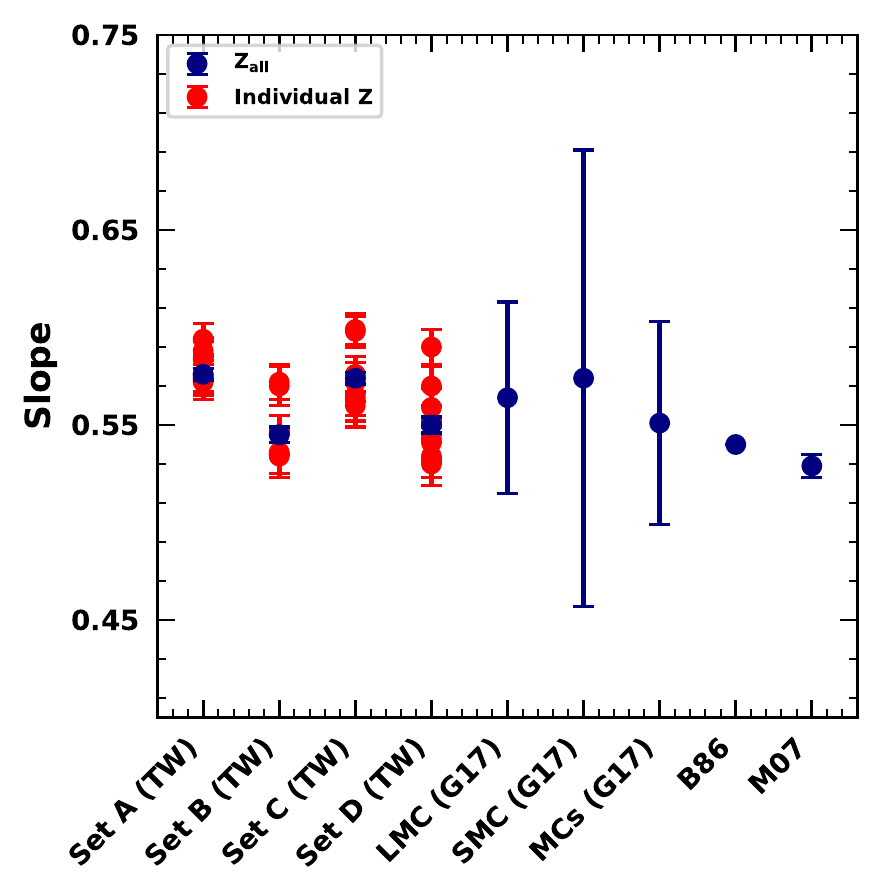}
\caption{Comparison of the slopes of the PR relations for the BL~Her
  stars obtained from this work (TW), using four different convective
  parameter sets, with those obtained from the literature. The red
  dots refer to the slopes of the PR relations obtained for the
  different chemical compositions individually, while the blue dots
  refer to the results obtained from considering the entire range of
  chemical compositions ($Z=0.00014$ to $Z=0.013$) combined. G17, B86
  and M07 refer to \cite{2017A&A...604A..29G},
  \cite{1986A&A...159..261B} and \cite{2007A&A...467..223M},
  respectively.}
\label{fig:PR}
\end{figure}

\section{Period--luminosity relations}

The mean magnitudes ($M_\lambda$) from the Fourier-fitted light curves
of the BL~Her models are used to obtain multi-wavelength PL relations
of the mathematical form,
\begin{equation}
M_\lambda = a\log(P)+b
\end{equation}
in a given band, $\lambda$. The multi-wavelength PL relations of the
BL~Her models computed using the four sets of convection parameters
are summarised in Table~\ref{tab:PL} and compared using the standard
$t$-test. In brief, we define a $T$ statistic for the comparison of
two linear regression slopes, $\hat{W}$, with sample sizes, $n$ and
$m$, as follows:
\begin{equation}
T=\frac{\hat{W}_n-\hat{W}_m}{\sqrt{\mathrm{Var}(\hat{W}_n)+\mathrm{Var}(\hat{W}_m)}},
\label{eq:ttest}
\end{equation}
where $\mathrm{Var}(\hat{W})$ is the variance of the slope. The null
hypothesis of equivalent slopes is rejected if $T>t_{\alpha/2,\nu}$ or
the probability of the observed value of the $T$ statistic is $p<0.05$
where $t_{\alpha/2,\nu}$ is the critical value under the two-tailed
$t$-distribution for the 95\% confidence limit ($\alpha$=0.05) and
degrees of freedom, $\nu=n+m-4$.

From Table~\ref{tab:PL} we find that the BL~Her models computed using
sets B and D (with radiative cooling) exhibit statistically similar PL
slopes at most wavelengths. In addition, although there exist
differences in the PL slopes of the BL~Her models computed using
different convection parameters, the empirical PL slopes of BL~Her
stars in the LMC seem to match well with the theoretical PL slopes at
most wavelengths. However, for a few particular cases ($HK_{\rm
  s}GG_{\rm BP}$), the empirical PL slopes of BL~Her stars in the LMC
seem to match better with the theoretical PL slopes obtained using
models with radiative cooling. Note here that the {\sl Gaia} apparent
magnitudes used to obtain the empirical PL relations have not been
corrected for extinction; however, to account for uncertainties
arising from extinction, we also obtain empirical PW relations using
Wesenheit magnitudes as defined by \citet{2019A&A...625A..14R} for
{\sl Gaia} DR3. We find the empirical PW slope of the BL~Her stars in
the LMC using {\sl Gaia} DR3 for calibration to be statistically
similar to the theoretical PW slopes computed using the four sets of
convection parameters.

The variation in the theoretical PL and PW relations obtained from the
BL~Her models as a function of metallicity is presented in
Figure~\ref{fig:PL_Zlambda}. We find strong effects of metallicity on
the theoretical PL relations in the $U$ and $B$ bands, possibly
because of increased sensitivity of bolometric corrections to
metallicity at wavelengths shorter than the $V$ band
\citep{2005oasp.book.....G, 2008ApJ...681..269K}. In the other bands,
the theoretical PL relations exhibit little to no metallicity
dependence; this is consistent with empirical PL relations from
\citet{2006MNRAS.370.1979M, 2009MNRAS.397..933M, 2011MNRAS.413..223M}
and \citet{2017A&A...604A..29G}. An interesting result to note is the
small but significant effect of metallicity on the theoretical PW
relations using the {\sl Gaia} passbands, although there seems to be
no effect of metallicity on the individual {\sl Gaia} passbands. We
will probe into the possible reasons of this metallicity dependence in
a future paper (Das et al., in prep.).

In addition, we also compare the PL relations obtained for the BL~Her
models with those from the recent grid of RR~Lyrae models from
\citet{2015ApJ...808...50M} in the $RIJHK_{\rm s}$ bands. From the
lowest panel of Table~\ref{tab:PL}, we find the PL slopes of the
RR~Lyrae models to be statistically similar in the $RIJK_{\rm s}$
bands to the BL~Her models computed without radiative cooling (sets~A
and C). This equivalence of the PL relations from RR~Lyrae and BL~Her
models is in agreement with empirical evidence from
\citet{2010JAVSO..38..100M}, \citet{2017AJ....153..154B} and
\citet{2020A&A...644A..95B} and with previous theoretical analysis
\citep[for example, see][]{ 2007A&A...471..893D, 2007A&A...467..223M}
and supports the claim that it could be useful to obtain common PL
relations from RR~Lyrae and Type~II Cepheids as an alternative to
classical Cepheids for the calibration of the extragalactic distance
scale.

\section{Period--radius relations}

We derive theoretical PR relations for the BL~Her models of the
mathematical form,
\begin{equation}
\log(R/{\rm R}_{\odot})=\alpha \log(P)+\beta,
\end{equation}
where $R$ is the mean radius of the BL~Her model obtained by averaging
the radius over a pulsation cycle.

A comparison of the slopes of the PR relations for the BL~Her models
computed using the four different convective parameter sets with those
obtained from previous results from \cite{1986A&A...159..261B},
\cite{2007A&A...467..223M} and \cite{2017A&A...604A..29G} is presented
in Figure~\ref{fig:PR}. We find that the BL~Her models computed
without radiative cooling (sets A and C) exhibit similar PR slopes
while those computed using radiative cooling (sets B and D) have
statistically similar PR slopes. The theoretical PR slopes of the
BL~Her models computed using the four sets of convection parameters
are in broad agreement with the PR slopes from earlier empirical and
theoretical results.

We also derive theoretical PRZ relations for the BL~Her models of the
form,
\begin{equation}
\log(R/{\rm R}_{\odot})=\alpha + \beta \log(P)+\gamma \mathrm{[Fe/H]}
\end{equation}
to test for the effects of metallicity on the PR relations. We find
the following relations for set A, the simplest convection parameter
set:
\begin{equation}
\begin{aligned}
\log(R/{\rm R}_{\odot})=&{}(0.879\pm 0.001)+(0.581\pm 0.003)\log(P)\\
&-(0.006\pm 0.001)\mathrm{[Fe/H]} \: (N=3266; \sigma=0.029),
\end{aligned}
\label{eq:PRZ-A}
\end{equation}
and for set D, the most complex convection parameter set:
\begin{equation}
\begin{aligned}
\log(R/{\rm R}_{\odot})=&{}(0.886\pm 0.002)+(0.554\pm 0.004)\log(P)\\
&-(0.006\pm 0.001)\mathrm{[Fe/H]} \: (N=2122; \sigma=0.029).
\end{aligned}
\label{eq:PRZ-D}
\end{equation}

The metallicity coefficient terms in Equations \ref{eq:PRZ-A} and
\ref{eq:PRZ-D} indicate a weak dependence of the PR relations on
metallicity, in agreement with previous empirical results from
\citet{1986A&A...159..261B} and \citet{2017A&A...604A..29G}.

\section{Summary and Conclusions}

We computed a very fine grid of BL~Her models, the shortest period
Type~II Cepheids using \textsc{mesa-rsp} covering a wide range of
input parameters: metallicity ($-2.0$ dex $\leq$ [Fe/H] $\leq$ 0.0
dex), stellar mass (0.5--0.8 M$_{\odot}$), luminosity (50--300
L$_{\odot}$) and effective temperature (full extent of the instability
strip; in steps of 50K) with non-linear pulsational periods typical of
BL~Her stars, i.e., $1\leq P ({\rm days}) \leq4$. The bolometric
luminosities that result as an output of the non-linear computations
were converted into absolute bolometric magnitudes and transformed
into absolute magnitudes in the respective passbands
(Johnson--Cousins--Glass $UBVRIJHKLL'M$ and {\sl Gaia} bands $GG_{\rm
  BP}G_{\rm RP}$) using pre-computed or user-provided bolometric
correction tables. The mean magnitudes obtained from Fourier fitting
the light curves were then used to derive theoretical PL and PW
relations.

As a function of different sets of convection parameters, the BL~Her
models computed using sets B and D (with radiative cooling) were found
to exhibit statistically similar PL slopes at most wavelengths. While
there exist differences among the theoretical PL relations of the
BL~Her models computed using the different sets of convection
parameters, the empirical PL relations of BL~Her stars in the LMC are
in broad agreement with their theoretical counterparts, with some
preference for models computed using radiative cooling especially in
the $HK_{\rm s}GG_{\rm BP}$ bands. The empirical PW slope of the
BL~Her stars in the LMC using {\sl Gaia} DR3 is statistically similar
to the theoretical PW slopes computed using the four sets of
convection parameters. As a function of metallicity, there exists a
strong effect of metallicity on the theoretical PL relations in the
$U$ and $B$ bands, with weak to negligible effects of metallicity at
longer wavelengths, which is in agreement with earlier empirical
results. We will probe the cause of the small but significant
metallicity effect on the theoretical PW relations using the {\sl
  Gaia} passbands in a future paper. 

We also found a weak dependence of the PR relations on metallicity. In
addition, the PL slopes of BL~Her models computed without radiative
cooling (sets~A and C) using \textsc{mesa-rsp} are statistically
similar to those for RR~Lyrae models from \citet{2015ApJ...808...50M}
in the $RIJK_{\rm s}$ bands, thereby supporting the claim that a
common PL relation using RR Lyrae and Type~II Cepheids could be used
as an alternative to classical Cepheids for extragalactic distance
scale calibrations. We also find the results from the
\textsc{mesa-rsp}-computed grid of BL~Her models to be in good
agreement with previous theoretical predictions from
\citet{2007A&A...471..893D, 2007A&A...467..223M} for BL~Her stars.

\section{Acknowledgements}
S. D.\ and L. M.\ acknowledge the KKP-137523 `SeismoLab' \'Elvonal grant
of the Hungarian Research, Development and Innovation Office
(NKFIH). M. J. gratefully acknowledges funding from MATISSE:
\textit{Measuring Ages Through Isochrones, Seismology, and Stellar
  Evolution}, awarded through the European Commission's Widening
Fellowship. This project has received funding from the European
Union's Horizon 2020 research and innovation program.


\begin{thebibliography}{}
\bibitem[Bhardwaj et al.(2017)]{2017AJ....153..154B} Bhardwaj, A., Macri, L.~M., Rejkuba, M., et al. \ 2017, \ AJ, 153, 154
\bibitem[Bhardwaj(2020)]{2020JApA...41...23B} Bhardwaj, A. \ 2020, \ JAA, 41, 23
\bibitem[Braga et al.(2020)]{2020A&A...644A..95B} Braga, V.~F., Bono, G., Fiorentino, G., et al. \ 2020, \ A\&A, 644, A95
\bibitem[Breuval et al.(2022)]{2022ApJ...939...89B} Breuval, L., Riess, A.~G., Kervella, P., Anderson, R.~I., \& Romaniello, M. \ 2022, \ ApJ, 939, 89
\bibitem[Burki and Meylan(1986)]{1986A&A...159..261B} Burki, G., \& Meylan, G. \ 1986, \ A\&A, 159, 261
\bibitem[Das et al.(2018)]{2018MNRAS.481.2000D} Das, S., Bhardwaj, A., Kanbur, S.~M., Singh, H.~P., \& Marconi, M. \ 2018, \ MNRAS, 481, 2000
\bibitem[Das et al.(2021)]{2021MNRAS.501..875D} Das, S., Kanbur, S.~M., Smolec, R., et al. \ 2021, \ MNRAS, 501, 875
\bibitem[De Somma et al.(2022)]{2022ApJS..262...25D} De Somma, G., Marconi, M., Molinaro, R., et al. \ 2022, \ ApJS, 262, 25
\bibitem[Di Criscienzo et al.(2007)]{2007A&A...471..893D} Di Criscienzo, M., Caputo, F., Marconi, M., \& Cassisi, S. \ 2007, \ A\&A, 471, 893
\bibitem[Gray(2005)]{2005oasp.book.....G} Gray, D.~F. \ 2005, \ The Observation and Analysis of Stellar Photospheres
\bibitem[Groenewegen and Jurkovic(2017)]{2017A&A...604A..29G} Groenewegen, M.~A.~T., \& Jurkovic, M.~I. \ 2017, \ A\&A, 604, A29
\bibitem[Kudritzki et al.(2008)]{2008ApJ...681..269K} Kudritzki, R.-P., Urbaneja, M.~A., Bresolin, F., et al. \ 2008, \ ApJ, 681, 269
\bibitem[Kuhfuss(1986)]{1986A&A...160..116K} Kuhfuss, R. \ 1986, \ A\&A, 160, 116
\bibitem[Lejeune et al.(1998)]{1998A&AS..130...65L} Lejeune, T., Cuisinier, F., \& Buser, R. \ 1998, \ A\&AS, 130, 65
\bibitem[Majaess(2010)]{2010JAVSO..38..100M} Majaess, D.~J. \ 2010, \ JAVSO, 38, 100
\bibitem[Marconi and Di Criscienzo(2007)]{2007A&A...467..223M} Marconi, M., \& Di Criscienzo, M. \ 2007, \ A\&A, 467, 223
\bibitem[Marconi et al.(2015)]{2015ApJ...808...50M} Marconi, M., Coppola, G., Bono, G., et al. \ 2015, \ ApJ, 808, 50
\bibitem[Matsunaga et al.(2006)]{2006MNRAS.370.1979M} Matsunaga, N., Fukushi, H., Nakada, Y., et al. \ 2006, \ MNRAS, 370, 1979
\bibitem[Matsunaga et al.(2009)]{2009MNRAS.397..933M} Matsunaga, N., Feast, M.~W., \& Menzies, J.~W. \ 2009, \ MNRAS, 397, 933
\bibitem[Matsunaga et al.(2011)]{2011MNRAS.413..223M} Matsunaga, N., Feast, M.~W., \& Soszy{\'n}ski, I. \ 2011, \ MNRAS, 413, 223
\bibitem[Paxton et al.(2019)]{2019ApJS..243...10P} Paxton, B., Smolec, R., Schwab, J., et al. \ 2019, \ ApJS, 243, 10
\bibitem[Ripepi et al.(2019)]{2019A&A...625A..14R} Ripepi, V., Molinaro, R., Musella, I., et al. \ 2019, \ A\&A, 625, A14
\bibitem[Ripepi et al.(2021)]{2021MNRAS.508.4047R} Ripepi, V., Catanzaro, G., Molinaro, R., et al. \ 2021, \ MNRAS, 508, 4047
\bibitem[Ripepi et al.(2022)]{2022A&A...659A.167R} Ripepi, V., Catanzaro, G., Clementini, G., et al. \ 2022, \ A\&A, 659, A167
\bibitem[Smolec and Moskalik(2008)]{2008AcA....58..193S} Smolec, R., \& Moskalik, P. \ 2008, \ AcA, 58, 193
\bibitem[Soszy{\'n}ski et al.(2008)]{2008AcA....58..293S} Soszy{\'n}ski, I., Udalski, A., Szyma{\'n}ski, M.~K., et al. \ 2008, \ AcA, 58, 293
\bibitem[Soszy{\'n}ski et al.(2018)]{2018AcA....68...89S} Soszy{\'n}ski, I., Udalski, A., Szyma{\'n}ski, M.~K., et al. \ 2018, \ AcA, 68, 89
\bibitem[Wallerstein(2002)]{2002PASP..114..689W} Wallerstein, G. \ 2002, \ PASP, 114, 689
\bibitem[Wielg{\'o}rski et al.(2022)]{2022ApJ...927...89W} Wielg{\'o}rski, P., Pietrzy{\'n}ski, G., Pilecki, B., et al. \ 2022, \ ApJ, 927, 89

\end{thebibliography}
\end{document}